\documentclass[pdflatex,sn-nature]{sn-jnl}

\usepackage{graphicx}%
\usepackage{multirow}%
\usepackage{amsmath,amssymb,amsfonts}%
\usepackage{amsthm}%
\usepackage{mathrsfs}%
\usepackage[title]{appendix}%
\usepackage{xcolor}%
\usepackage{textcomp}%
\usepackage{manyfoot}%
\usepackage{booktabs}%
\usepackage{algorithm}%
\usepackage{algorithmicx}%
\usepackage{algpseudocode}%
\usepackage{listings}%

\theoremstyle{thmstyleone}%
\theoremstyle{thmstyletwo}%
\theoremstyle{thmstylethree}%
\begin{document}

\title[Article Title]{Hole-doped superconductivity above 100 K in infinite-layer cuprate thin films}

\author*[1]{\fnm{Biemeng} \sur{Jin}}\email{biemeng.jin@u.nus.edu}
\equalcont{These authors contributed equally to this work.}
\author[1]{\fnm{Saurav}    \sur{Prakash}}
\equalcont{These authors contributed equally to this work.}
\author[1]{\fnm{Zhaoyang}  \sur{Luo}}
\author[2]{\fnm{Shengwei}  \sur{Zeng}}
\author[3,4]{\fnm{Jing-Yang}  \sur{Chung}}  
\author[1]{\fnm{Xing}      \sur{Gao}}
\author[2]{\fnm{Zhi Shiuh} \sur{Lim}}
\author[1]{\fnm{Jiangbo}    \sur{Luo}}
\author[1]{\fnm{King} \spfx{Yau} \sur{Yip}}
\author[1]{\fnm{Wei}        \sur{Zhang}}
\author[1]{\fnm{Nurul}     \sur{Fitriyah}}
\author[1]{\fnm{Shuhan}    \sur{Lu}}
\author[1]{\fnm{Taiyu}    \sur{An}}
\author[5]{\fnm{Ping}       \sur{Yang}}
\author[3]{\fnm{Qian}  \sur{He}}  
\author[3,4]{\fnm{Silvija}  \sur{Gradečak}}
\author[2]{\fnm{Huajun}  \sur{Liu}}
\author*[1]{\fnm{A.} \sur{Ariando}}\email{ariando@nus.edu.sg}

\affil*[1]{\orgdiv{Department of Physics}, \orgdiv{Faculty of Science}, \orgname{National University of Singapore}, \orgaddress{\city{Singapore}, \postcode{117551}, \country{Singapore}}}

\affil[2]{\orgdiv{Institute of Materials Research and Engineering (IMRE)}, \orgdiv{Agency for Science}, \orgname{Technology and Research (A*STAR)}, \orgaddress{\city{Singapore}, \postcode{138634}, \country{Singapore}}}

\affil[3]{\orgdiv{Department of Materials Science and Engineering}, \orgname{National University of Singapore}, \orgaddress{\city{Singapore}, \postcode{117575}, \country{Singapore}}}

\affil[4]{\orgdiv{Applied Materials—NUS Advanced Materials Corporate Lab}, \orgaddress{\city{Singapore}, \postcode{117608}, \country{Singapore}}}

\affil[5]{\orgdiv{Singapore Synchrotron Light Source (SSLS)}, \orgname{National University of Singapore}, \orgaddress{\city{Singapore}, \postcode{117603}, \country{Singapore}}}

\abstract{ 
Since the discovery of superconductivity in (La,Ba)$_2$CuO$_2$ (Ref.~\cite{bednorz1986possible}), a broad family of structurally distinct cuprate superconductors has been proposed or engineered to elucidate the physics of high-temperature superconductivity~\cite{chu2015hole,plakida2010high}. Among them, the infinite-layer cuprate has the simplest structure, consisting only of the essential ingredients for superconductivity: CuO$_2$ square planes separated by spacer ions~\cite{siegrist1988parent}. Despite being proposed nearly 40 years ago, the hole-doped superconductivity via chemical substitution in this compound has not yet been achieved, a fundamental open question in the field. Here, we report the observation of superconductivity in the hole-doped infinite-layer cuprate thin film. Measurements of resistivity and magnetic-field response in Sr$_{1-x}$Rb$_x$CuO$_2$ single-crystal thin films show superconducting transitions with a high onset temperature of 100 K. Hole doping is achieved via the synergistic effect of rubidium substitution and apical oxygen incorporation, as evidenced by structural analysis and transport measurements. As the parent structure of the cuprate family~\cite{chu2015hole}, hole-doped infinite-layer cuprate provides a unique platform for revisiting key puzzles in cuprate superconductors~\cite{keimer2015quantum,tsuei2000pairing,armitage2010progress,dagotto1994correlated}, including strange metal~\cite{proust2019remarkable,taillefer2010scattering} and electron-hole symmetry~\cite{tohyama2004asymmetry,segawa2010zero,lee2014asymmetry}, while bridging to cuprate-nickelate symmetry~\cite{li2019superconductivity,zeng2022superconductivity,chow2025bulk,lechermann2020late}.
}

\maketitle

\section*{Main}
\label{sec1}
\begin{figure*}[h]
\centering
\includegraphics[width=0.9\columnwidth]{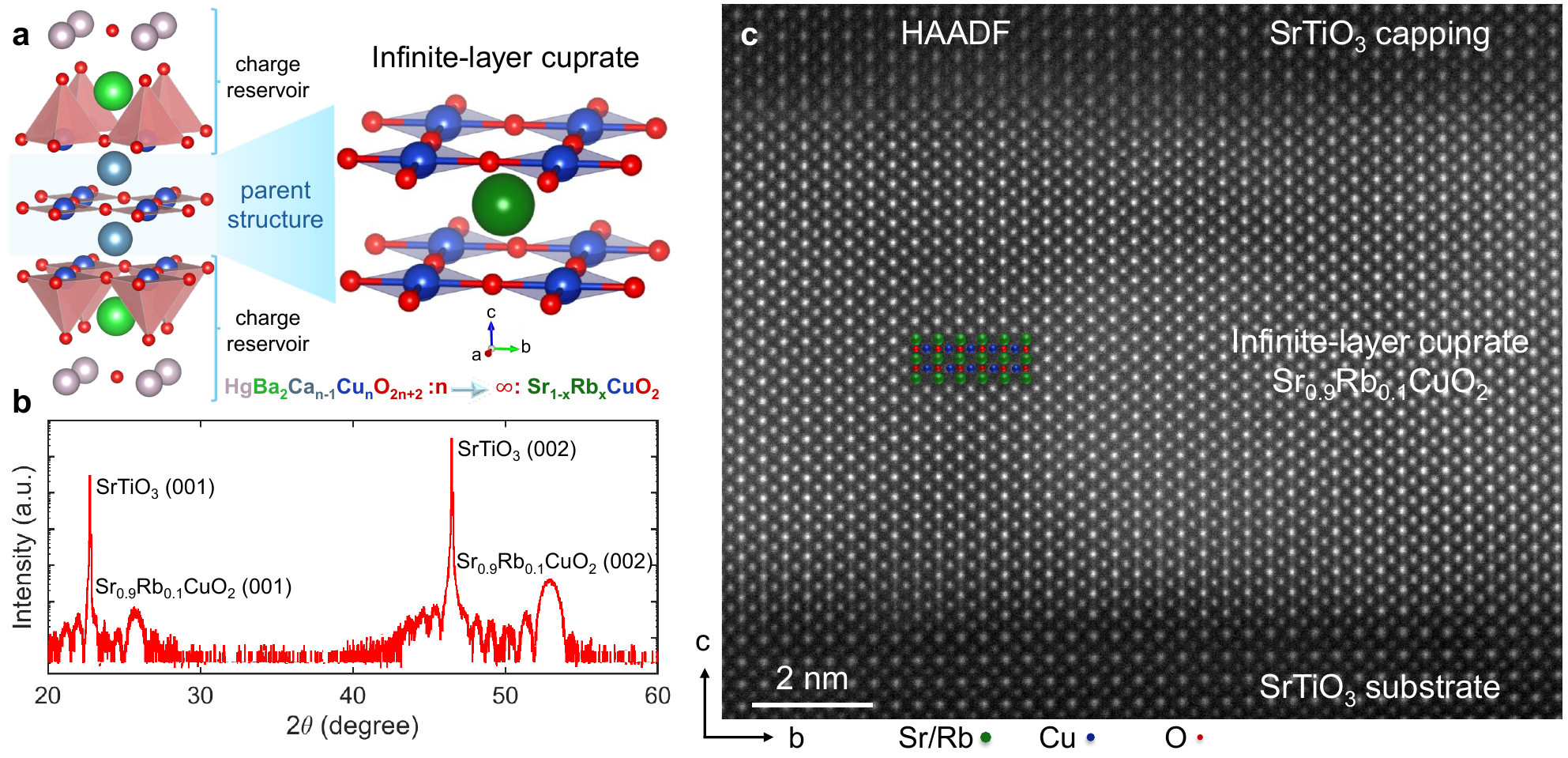}
\caption{
\textbf{Structural characterization of infinite-layer cuprate thin films $\mathrm{Sr_{1-x}Rb_xCuO_2}$.} 
\textbf{a,} Schematic crystal structure of infinite-layer cuprate $\mathrm{Sr_{1-x}Rb_xCuO_2}$ (right), representing the simplest cuprate structure which is derived from the infinite-layer limit of $\mathrm{HgBa_2Ca_{n-1}Cu_nO_{2n+2}}$ (left) by removing the complex charge-reservoir blocks.
\textbf{b,} X-ray diffraction symmetric $\theta$–$2\theta$ scans of $\mathrm{Sr_{0.9}Rb_{0.1}CuO_2}$ thin films grown on $\mathrm{SrTiO_3}$ (001) substrates, showing clear (001) and (002) peaks of the infinite-layer cuprate.
\textbf{c,} Cross-sectional STEM high-angle annular dark-field (HAADF) image showing the well-defined infinite-layer structure. a.u., arbitrary units.
}\label{fig1}
\end{figure*}

Cuprate superconductors (such as $\mathrm{HgBa_2Ca_{n-1}Cu_nO_{2n+2}}$) typically feature complex structures with large charge-reservoir blocks~\cite{chu2015hole,dagotto1994correlated}, which complicates the understanding of the key physics underlying high-temperature superconductivity. This motivates the search for simpler structures. By pushing $\mathrm{HgBa_2Ca_{n-1}Cu_nO_{2n+2}}$ to its infinite-layer limit, the infinite-layer cuprate $\mathrm{ACuO}_2$ (Fig.~\ref{fig1}a) has been proposed as the simplest and most minimal structure among all cuprates~\cite{siegrist1988parent}, containing only $\mathrm{CuO}_2$ square planes separated by alkaline-earth ions $\mathrm{A}$. Despite its structural simplicity, stabilizing the infinite-layer cuprate is very challenging. The first synthesis of this cuprate was reported by T. Siegrist in 1988~\cite{siegrist1988parent}, but it was not superconducting. Shortly thereafter, using high-temperature and high-pressure solid-state synthesis, superconductivity was observed in the electron-doped phase~\cite{smith1991electron} and in defective infinite-layer cuprate~\cite{azuma1992superconductivity}, the latter of which is believed to host hole carriers introduced by planar defect layers. Meanwhile, $\mathrm{SrCuO_2}/\mathrm{BaCuO_2}$ superlattice was also found to exhibit superconductivity~\cite{norton1994superconductivity}. In contrast, despite decades of attempts~\cite{kubo1994synthesis,li1994diamagnetic,qin2005synthesis,yakabe1994p,hadjimichael2022structural,waelchli2021thesis}, hole-doped superconducting infinite-layer cuprate via chemical substitution—the most sought-after case for understanding the mechanism of high-$T_c$ superconductivity—remain elusive.

Over the past 40 years, intensive efforts have been devoted to realizing superconductivity in hole-doped infinite-layer cuprate $\mathrm{ACuO}_2$ via chemical substitution. On the $\mathrm{A^{2+}}$ site, small monovalent cations ($\mathrm{Li^+}$ and $\mathrm{Na^+}$) were doped into infinite-layer cuprate~\cite{kubo1994synthesis,li1994diamagnetic,qin2005synthesis,yakabe1994p}, while on the oxygen site, excess oxygen was incorporated via oxygen-rich growth techniques~\cite{yakabe1994p,hadjimichael2022structural,waelchli2021thesis}. However, both approaches resulted in insulating behavior. Decades of attempts have been unsuccessful hitherto. Here, we propose to synergistically combine rubidium doping and oxygen incorporation to achieve hole-doped superconductivity in infinite-layer cuprate $\mathrm{Sr_{1-x}Rb_xCuO_2}$ for the following considerations (despite slight off-stoichiometry of rubidium and oxygen, the stoichiometric formula is used in this manuscript for simplicity). First, rubidium, as a large dopant, may mitigate the limitations of smaller dopants~\cite{shibata1990superconductivity,kubo1994synthesis,li1994diamagnetic,qin2005synthesis,yakabe1994p} (including volatile loss and hole compensation by dopant-induced oxygen vacancies). The effectiveness of rubidium is further evidenced by bond valence sum analysis~\cite{yamauchi1998hole,waelchli2021thesis}, which indicates that larger dopants provide hole carriers more efficiently, and by empirical observations~\cite{waelchli2021thesis,weber2012scaling} that carrier delocalization induced by large cations favors superconductivity. Secondly, slow-growth pulsed laser deposition followed by oxygen-rich annealing (see Methods) is optimized to incorporate extra oxygen, reducing oxygen vacancies and introducing apical oxygen for hole doping.

Through systematic optimization of growth conditions, rubidium-doped cuprate exhibits a superior infinite-layer structure, as demonstrated by X-ray diffraction (XRD) and scanning transmission electron microscopy (STEM). Figure~\ref{fig1}b shows a representative XRD symmetric $\theta$-$2\theta$ scan of $\mathrm{Sr_{0.9}Rb_{0.1}CuO_2}$ films grown on $\mathrm{SrTiO_3}$ substrates. Only the (001) and (002) peaks of the infinite-layer cuprate are observed, in agreement with previous reports~\cite{waelchli2021thesis,hadjimichael2022structural,takano1989acuo2}, which confirms that the samples have the well-defined infinite-layer structure. The pronounced Laue oscillations around these peaks further indicate the high-quality crystallinity of the films. The out-of-plane lattice constant, calculated from the (002) peak ($2\theta \approx 52.9^\circ$), is about 3.46~\AA, slightly larger than the undoped bulk value of 3.43~\AA~\cite{takano1989acuo2}. The in-plane lattice constant is approximately 3.91~\AA, as determined from reciprocal space mapping ($Q_{[100]}\approx 1.6~\mathrm{\AA^{-1}}$ from Extended Data Fig.~\ref{extendeddatafig1}a). This indicates that the infinite-layer cuprate film is under compressive strain from the $\mathrm{SrTiO_3}$ substrate ($\sim$0.8\%). High-angle annular dark-field (HAADF) images in Figure~\ref{fig1}c and Extended Data Figure~\ref{extendeddatafig1}(b-c) further confirm the good infinite-layer structure without impurity phases in the film.

\begin{figure*}[h]
\centering
\includegraphics[width=0.9\columnwidth]{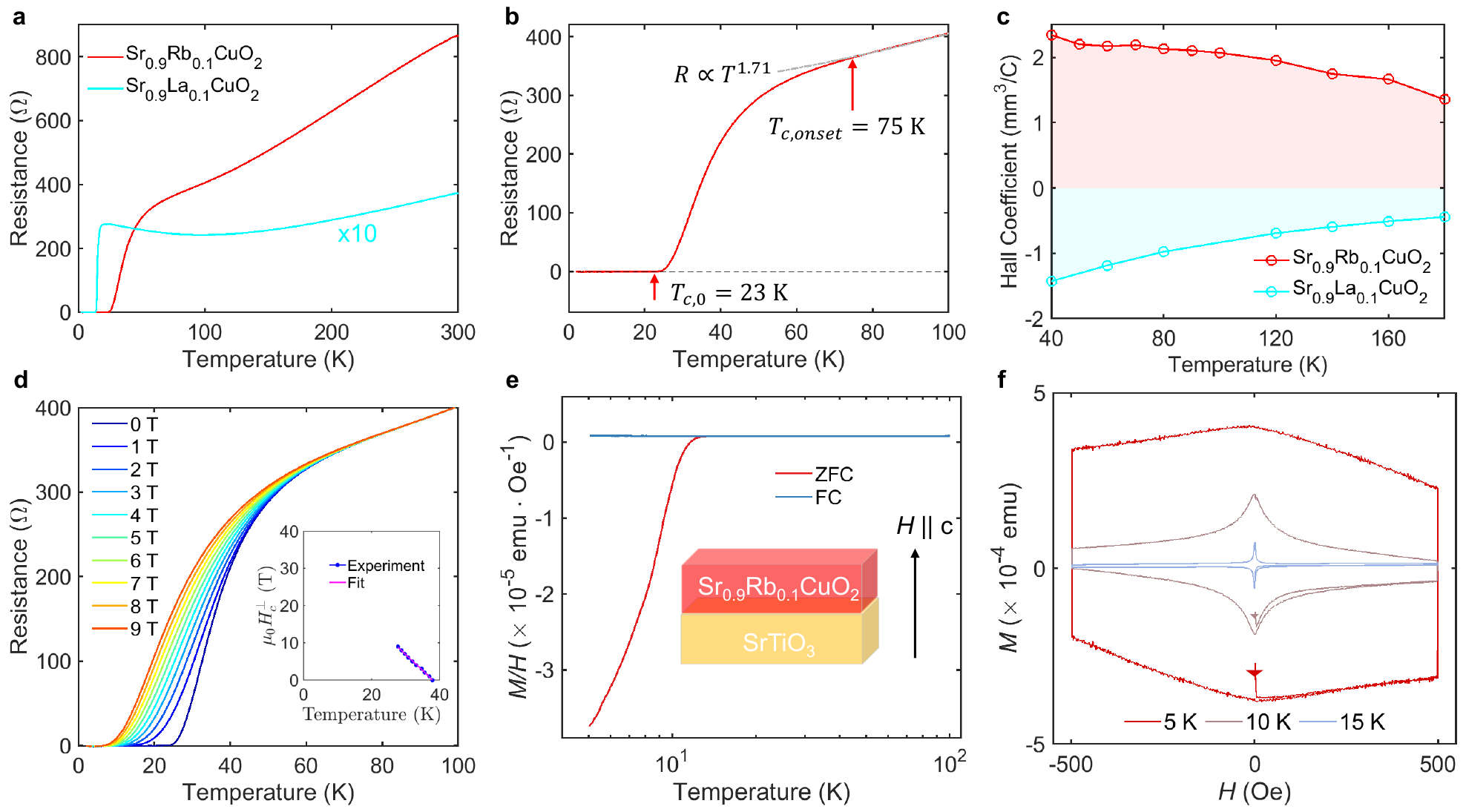}
\caption{
\textbf{Observation of hole-doped superconductivity in  infinite-layer cuprate $\mathrm{Sr_{1-x}Rb_xCuO_2}$.}
\textbf{a,} Temperature dependence of resistance $R(T)$ for hole-doped $\mathrm{Sr_{0.9}Rb_{0.1}CuO_2}$ (red) and electron-doped $\mathrm{Sr_{0.9}La_{0.1}CuO_2}$ (cyan) infinite-layer cuprates. The resistance of the electron-doped sample is multiplied by 10 for clarity. The hole-doped infinite-layer cuprate exhibits linear-in-temperature behavior at high temperatures and has a much higher critical temperature compared to its electron-doped counterpart.
\textbf{b,} Determination of the critical temperature for a representative superconducting $\mathrm{Sr_{0.9}Rb_{0.1}CuO_2}$ thin film. The onset temperature ($T_{c,onset}$) is approximately 75~K, defined by the deviation from a power-law fit $R \propto T^{\alpha}$ in the range 110-170~K (dashed line). The zero-resistance temperature ($T_{c,0}$) is about 23~K, where the resistance becomes indistinguishable from the noise. The broad superconducting transition likely reflects a limited superconducting volume fraction.
\textbf{c,} Temperature dependence of the Hall coefficient for $\mathrm{Sr_{0.9}Rb_{0.1}CuO_2}$ and $\mathrm{Sr_{0.9}La_{0.1}CuO_2}$, confirming hole-type carriers in the Rb-doped samples in contrast to the well-established electron-doped counterpart.
\textbf{d,} Magnetic-field dependence of resistance up to 9~T for superconducting $\mathrm{Sr_{0.9}Rb_{0.1}CuO_2}$ with the field applied perpendicular to the $ab$ planes. The inset shows the upper critical field $H_{c}^{\perp}$ (blue), defined at the midpoint of the resistive transition, with a linear fit (magenta) near $T_{c}$.
\textbf{e,} Direct-current magnetic susceptibility of a $\mathrm{Sr_{0.9}Rb_{0.1}CuO_2}$ thin film measured under a magnetic field perpendicular to the $ab$ planes in zero-field cooling (ZFC, red) and field cooling (FC, blue). A clear diamagnetic response appears below $\sim$13~K, lower than $T_{c,onset}$, consistent with a limited superconducting volume fraction.
\textbf{f,} Magnetization–field ($M$–$H$) loops measured at 5~K, 10~K, and 15~K, showing magnetic flux exclusion at low fields (arrows) and hysteresis characteristic of the Meissner effect.
}\label{fig2}
\end{figure*}

Figure~\ref{fig2}a presents the temperature-dependent resistance of $\mathrm{Sr_{0.9}Rb_{0.1}CuO_2}$ infinite-layer cuprate thin films, along with that of the well-established electron-doped counterpart $\mathrm{Sr_{0.9}La_{0.1}CuO_2}$. As temperature decreases, $\mathrm{Sr_{0.9}Rb_{0.1}CuO_2}$ exhibits a linear-in-temperature resistivity at high temperatures, followed by an upward curvature, and clearly enters the superconducting state at low temperature. The superconducting onset temperature $T_{c,onset}$ is approximately 75~K, determined from the deviation of the resistance from a power-law fit $R \propto T^{\alpha}$ in the 110-170~K range (Fig.~\ref{fig2}b). This onset temperature is significantly higher than that of electron-doped infinite-layer cuprate (the reported maximum $T_{c}$ for the latter is about 44~K~\cite{smith1991electron}). The zero-resistance temperature $T_{c,0}$ is approximately 23~K. The broad superconducting transition likely arises from a limited superconducting volume within the film.
To probe the nature of charge carriers in this new superconducting system, Hall measurements were performed, as shown in Figure~\ref{fig2}c. In contrast to the well-known electron-doped $\mathrm{Sr_{0.9}La_{0.1}CuO_2}$, the Hall coefficient of $\mathrm{Sr_{0.9}Rb_{0.1}CuO_2}$ remains positive and increases upon decreasing from 180~K to 40~K, indicating hole-type charge carriers. This confirms hole-doped superconductivity in $\mathrm{Sr_{0.9}Rb_{0.1}CuO_2}$ infinite-layer cuprate and suggests predominantly single-band transport, consistent with other cuprate members. Overall, the long-sought superconductivity in hole-doped infinite-layer cuprate has been successfully realized in $\mathrm{Sr_{0.9}Rb_{0.1}CuO_2}$ thin films and is highly reproducible (see Extended Data Fig.~\ref{extendeddatafig2}a), providing a unique platform to address key puzzles in cuprate superconductors, such as electron-hole symmetry, strange metal and the role of charge reservoir blocks.

Together with the robust zero-resistance state, temperature-dependent magnetoresistance (Fig.~\ref{fig2}d) and the observation of the Meissner effect (Fig.~\ref{fig2}e,f) unambiguously confirm superconductivity in hole-doped $\mathrm{Sr_{0.9}Rb_{0.1}CuO_2}$ infinite-layer cuprate thin films. Figure~\ref{fig2}d presents the magnetoresistance measured under magnetic fields up to 9~T applied perpendicular to the $ab$ planes. The superconducting transition is progressively suppressed with increasing magnetic field below the onset temperature $T_{c,onset} \approx$75~K, consistent with the previous determination. To estimate the perpendicular upper critical field $H_{c}^{\perp}$, the midpoint of the resistive transition near $T_{c,onset}$ is taken as the criterion (inset of Fig.~\ref{fig2}d). The extracted data are fitted using the linearized Ginzburg–Landau relation
\begin{equation}
H_{c}^{\perp} = \frac{\Phi_0}{2\pi \xi_{GL}^2(0)}\left(1-\frac{T}{T_c}\right),
\end{equation}
where $\Phi_0$ is the magnetic flux quantum and $\xi_{GL}(0)$ is the Ginzburg–Landau coherence length at zero temperature. The fit yields $H_{c}^{\perp}(0) \approx 33$~T and $\xi_{GL}(0) \approx 3$~nm, values comparable to those of other cuprate superconductors~\cite{proust2019remarkable}. The magnetic properties of the superconducting $\mathrm{Sr_{0.9}Rb_{0.1}CuO_2}$ infinite-layer cuprate thin films are further characterized. Figure~\ref{fig2}e shows the direct-current magnetic susceptibility of $\mathrm{Sr_{0.9}Rb_{0.1}CuO_2}$ films measured with the magnetic field perpendicular to the $ab$ planes under zero-field cooling (ZFC) and field cooling (FC) conditions. A strong diamagnetic response is clearly observed below $\sim$13~K. (This value is lower than $T_{{c,onset}} = 75$~K, likely due to a limited superconducting volume~\cite{di2012occurrence,li2019superconductivity,chow2025bulk,zhou2025ambient}.) The small positive signal above 13~K originates from the background like $\mathrm{SrTiO_3}$ substrates. Furthermore, magnetization–field ($M$–$H$) loops measured at 5~K, 10~K, and 15~K (Fig.~\ref{fig2}f) exhibit magnetic flux exclusion (see the arrows) and clear hysteresis loop, characteristic of the Meissner state.

\begin{figure*}[h]
\centering
\includegraphics[width=0.9\columnwidth]{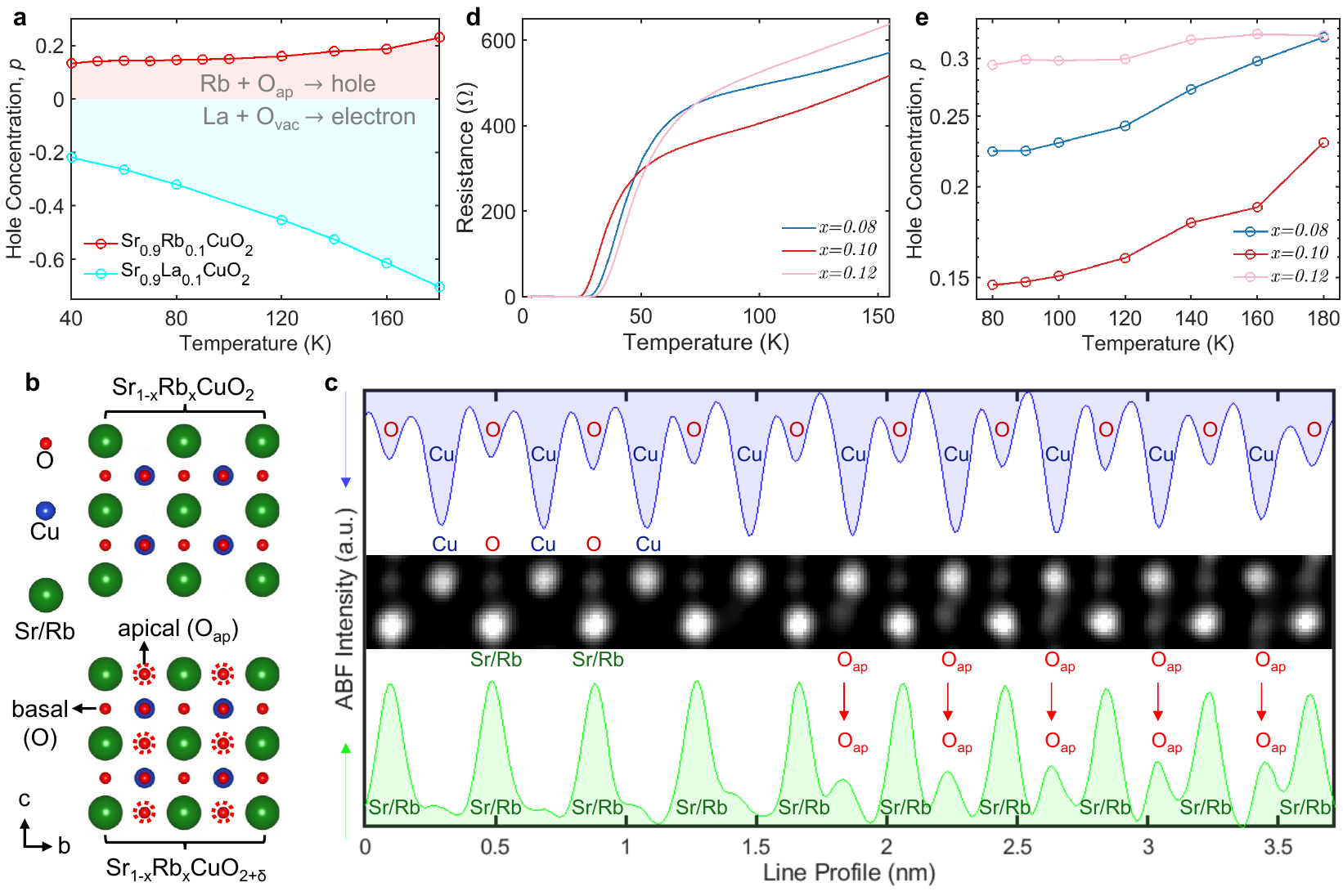}
\caption{
\textbf{Synergistic hole doping via rubidium substitution and apical oxygen incorporation in infinite-layer cuprate $\mathrm{Sr_{0.9}Rb_{0.1}CuO_2}$.}
\textbf{a,} Hole concentration of hole-doped $\mathrm{Sr_{0.9}Rb_{0.1}CuO_2}$ and electron-doped $\mathrm{Sr_{0.9}La_{0.1}CuO_2}$ infinite-layer cuprate, estimated from Hall measurements (Fig.~\ref{fig2}c). The hole concentration for both compounds is exceeding the nominal $x =$ 0.10 introduced purely by heterovalent substitution ($p\approx 0.15$ for $\mathrm{Rb^+}$ doping at 80~K, and $p\approx 0.22$ for $\mathrm{La^{3+}}$ doping at 40~K). In hole-doped $\mathrm{Sr_{0.9}Rb_{0.1}CuO_2}$, additional hole carriers arise from apical oxygen incorporation enabled by oxygen-rich annealing, whereas in electron-doped $\mathrm{Sr_{0.9}La_{0.1}CuO_2}$, extra electron carriers originate from oxygen vacancies induced by vacuum annealing.
\textbf{b,} Supercell structural model of $\mathrm{Sr_{1-x}Rb_xCuO_{2+\delta}}$, illustrating the spatial distribution of apical oxygen ($\mathrm{O_{ap}}$) and basal ($\mathrm{O}$) oxygen sites within the lattice. Red dashed circles denote partially occupied apical oxygen sites.
\textbf{c,} Intensity line profiles extracted from the spacer layer (bottom) and CuO$2$ layer (top), as indicated in the inset of the noise-filtered STEM-ABF image (average background subtraction filter). Bright features correspond to atomic columns: Cu and $\mathrm{O}$ in the CuO$2$ layer, and Sr/Rb and  $\mathrm{O_{ap}}$ in the spacer layer. The corresponding raw image is provided in Extended Data Fig.~\ref{extendeddatafig3}. The doublet peaks in the spacer layer profile indicates the presence of apical oxygen  ($\mathrm{O_{ap}}$).
\textbf{d,} Temperature-dependent resistance of hole-doped infinite-layer cuprates $\mathrm{Sr_{1-x}Rb_xCuO_2}$ with doping ratio $x = 0.08$, 0.10, and 0.12.
\textbf{e,} Hole concentration of $\mathrm{Sr_{1-x}Rb_xCuO_2}$ with different $x =$ 0.08, 0.10, and 0.12, calculated from Hall Measurements (Extended Data Fig.~\ref{extendeddatafig2}c). 
The $x = 0.12$ sample exhibits higher critical temperatures (both onset and zero-resistance) and a greater hole concentration than the $x = 0.10$ sample, consistent with enhanced hole doping from increased Rb substitution. Similarly, the $x = 0.08$ sample also shows higher critical temperatures and hole concentration compared to the $x = 0.10$ sample, possibly related to enhanced apical oxygen incorporation (less rubidium substitution introduces less oxygen vacancies, leaving more oxygens to occupy the apical sites for hole doping).
}\label{fig3}
\end{figure*}

Hole doping in the infinite-layer cuprate $\mathrm{SrCuO_2}$ can typically be realized from two degrees of freedom: substitution of $\mathrm{Sr^{2+}}$ by monovalent cations at the Sr site, and incorporation of apical oxygen at the oxygen site. For $\mathrm{Sr_{0.9}Rb_{0.1}CuO_2}$, Rb substitution ($x = 0.10$) is expected to yield a theoretical hole concentration of $p = 0.10$. (Although Fermi surface reconstruction can give rise to a larger hole concentration, this typically occurs only at higher doping ($x > 0.2$)~\cite{pelc2020resistivity,proust2019remarkable}, which is well above the present case of $x = 0.10$.) However, Figure~\ref{fig3}a clearly shows that the measured hole concentration is higher than $p \approx 0.10$ prior to the onset of superconductivity. This suggests the presence of extra oxygen in $\mathrm{Sr_{0.9}Rb_{0.1}CuO_2}$ that contributes to additional hole carriers. Thus, it is the synergistic combination of Rb substitution and apical oxygen incorporation that introduces hole carriers in the infinite-layer cuprate $\mathrm{Sr_{1-x}Rb_xCuO_2}$. This is corroborated as follows. First, the oxygen sites in the infinite-layer cuprate are active, capable of incorporating apical oxygen or forming oxygen vacancies. In hole-doped $\mathrm{Sr_{0.9}Rb_{0.1}CuO_2}$, extra oxygen is likely incorporated at apical sites during film growth and oxygen-rich post-annealing, providing additional hole carriers. This activity at the oxygen sites is also observed in the electron-doped counterpart~\cite{smith1991electron,armitage2010progress,yan2025enhancement}. For example, in $\mathrm{Sr_{0.9}La_{0.1}CuO_2}$, the measured electron concentration exceeds the nominal La substitution level ($p > x = 0.1$ from Fig.~\ref{fig3}a), with extra electron carriers introduced by oxygen vacancies during vacuum annealing~\cite{armitage2010progress}. Second, apical oxygen illustrated in the supercell structure of $\mathrm{Sr_{1-x}Rb_xCuO_{2+\delta}}$ (Fig.~\ref{fig3}b) is experimentally observed. Figure~\ref{fig3}c shows line profiles of the spacer (bottom) and CuO$_2$ (top) layers extracted from STEM-ABF images. In the spacer layer, apart from single peaks corresponding to Sr or Rb atoms, doublet peaks are occasionally observed at the apical sites. In fact, these doublet peaks correspond to apical oxygen ($\mathrm{O_{ap}}$) within the spacer layer. Third, synergistic hole-doping is further hinted by doping dependence. The samples for both higher ($x = 0.12$) and lower ($x = 0.08$) rubidium doping relative to $x = 0.10$ exhibit higher critical temperatures $T_c$ and larger hole concentrations $p$, as shown in Figure~\ref{fig3}d,e. For $x = 0.12$, increased Rb substitution provides more hole carriers, resulting in higher $T_c$ and $p$ (all three doping levels remain in the underdoped regime as their hole concentrations are low~\cite{pelc2020resistivity}). In contrast, the $x = 0.08$ sample also shows enhanced $T_c$ and $p$ despite reduced Rb content. This may arise because lower Rb substitution introduces fewer oxygen vacancies~\cite{shibata1990superconductivity}, leaving more oxygens at apical sites. Consequently, such synergetic doping produces more hole carriers than in the $x = 0.10$ sample. This resultant hole-doping from Rb substitution, apical oxygen, and oxygen vacancies induced by Rb may lead to a non-trivial superconducting dome, which will be explored in future studies.

In summary, we have achieved hole-doped superconductivity in infinite-layer cuprate thin films, a long-standing experimental challenge since its proposal nearly four decades ago~\cite{siegrist1988parent,chu2015hole}. This new system provides a unique platform for understanding the fundamental mechanism of high-temperature superconductivity. First, as the parent structural motif of all cuprates, the hole-doped infinite-layer compound addresses key puzzles in cuprate physics~\cite{keimer2015quantum,tsuei2000pairing,proust2019remarkable,taillefer2010scattering,armitage2010progress,dagotto1994correlated}. On the one hand, its minimal architecture unequivocally demonstrates that charge-reservoir blocks are not necessary for the emergence of superconductivity in curpates. On the other hand, the observed linear-in-temperature resistivity (Fig.~\ref{fig2}a) may suggest strange-metal behavior in this new system, which would offer a new avenue for studying its underlying physics. Second, the highest superconducting transition temperature among cuprates (133~K reported in $\mathrm{HgBa_2Ca_2Cu_3O_{8+\delta}}$ from Ref.~\cite{schilling1993superconductivity}) remains unsurpassed. The realization of hole-doped superconductivity in hole-doped infinite-layer cuprates with an onset temperature around 100~K opens new possibilities for further enhancing superconducting $T_{c}$. In particular, tuning coupling and tailoring spin fluctuations via ionic-radius engineering may provide viable routes to elevating $T_{c}$—strategies reminiscent of recent advances in nickelate superconductors~\cite{li2019superconductivity,zeng2022superconductivity,chow2025bulk,sun2023signatures,zhou2025ambient}. Finally, as the isostructural analog of recently-discovered superconducting infinite-layer nickelates~\cite{li2019superconductivity,zeng2022superconductivity,chow2025bulk}, hole-doped infinite-layer cuprates are pivotal in bridging the physics of cuprate and nickelate superconductors, offering a promising pathway toward a unified understanding of high-temperature superconductivity. For example, understanding electron–hole symmetry in infinite-layer cuprates may shed light on the so-far absent electron-doped superconductivity in nickelates.

\backmatter

\bibliography{reference}

\bmhead{Data Availability}
The data that support the findings of this study are available from the corresponding authors upon reasonable request.

\section*{Method}
\label{method}
\bmhead{Growth of $\mathrm{Sr_{1-x}Rb_{x}CuO_2}$ thin films}
Ceramic targets were prepared by a conventional solid-state reaction using high-purity  $\mathrm{SrCO_3}$ (99.99\%, Alfa-Aesar), $\mathrm{Rb_2CO_3}$ (99.8\%, Sigma-Aldrich), and $\mathrm{CuO}$ (99.9995\%,  Alfa-Aesar) powders (weighted according to the chemical formula of $\mathrm{Sr_{1-x}Rb_{x}CuO_2}$). To ensure complete decarbonation, the mixed powders were sintered in air for 12~h at 900~$^\circ$C and 950~$^\circ$C, respectively, with thorough regrinding before each sintering. The resulting powder was then pressed into disk-shaped pellets and sintered at 1000~$^\circ$C for 15~h. HF-treated $\mathrm{SrTiO_3}$(001) substrates (HFKJ) were annealed in air at 1000~$^\circ$C for 1.5~h to obtain atomically flat surfaces with well-defined step-and-terrace surface. $\mathrm{Sr_{1-x}Rb_{x}CuO_2}$ infinite-layer thin films were deposited on $\mathrm{TiO_2}$-terminated $\mathrm{SrTiO_3}$(001) substrates using pulsed laser deposition (PLD) at a temperature of 550~$^\circ$C and an oxygen partial pressure of 150~mTorr. A 248~nm KrF excimer laser with a deposition rate of 1~Hz was used, and the laser energy density on the target surface was approximately 2.0~J\,cm$^{-2}$. Immediately after deposition, the films were capped with a 4~nm-thick $\mathrm{SrTiO_3}$ layer under the same growth conditions to prevent degradation of the hole-doped infinite-layer cuprates. The laser energy density for the capping layer was approximately 1.4~J\,cm$^{-2}$. After deposition, the films were cooled to 500~$^\circ$C for an in-situ oxygen-rich annealing lasting 45~min under a high oxygen pressure of 200~Torr to incorporate oxygen and reduce possible oxygen vacancies. Finally, the samples were cooled to room temperature.

\bmhead{X-ray diffraction (XRD)}
High-resolution X-ray diffraction (XRD) $\theta$–$2\theta$ symmetric scans and reciprocal space mapping (RSM) were performed using a XRD facility (SmartLab, Rigaku) with an X-ray wavelength of $\lambda = 1.5404$~\AA\ at the Institute of Materials Research and Engineering (IMRE), Agency for Science, Technology and Research (A*STAR). Additional measurements were carried out at the X-ray Diffraction and Development beamline of the Singapore Synchrotron Light Source (SSLS) using an X-ray wavelength of $\lambda = 1.5404$~\AA.

\bmhead{Scanning Transmission Electron Microscopy (STEM)}
To prepare cross-sectional specimens, the thin-film samples were first capped with a protective layer of gold or platinum. A focused ion beam (FIB) instrument (FEI Versa 3D) was employed for lamella preparation, with the operating voltage decreasing from 30 to 5~kV and tilting angles ranging from $\pm 1.5^\circ$ to $\pm 3.5^\circ$. Final polishing was performed at 2~kV with a tilting angle of $\pm 6^\circ$ to remove the amorphous surface layer. STEM characterization was conducted on a JEOL JEM-ARM200F microscope operated at 200~kV, equipped with a cold field emission gun and a probe aberration corrector. HAADF and ABF images were acquired simultaneously with a convergence semi-angle of about 30~mrad. The collection semi-angles were optimized at about 70-280~mrad for HAADF imaging to capture high-Z contrast, and about 15-30~mrad for ABF imaging to resolve the oxygen columns. All imaging was performed with the electron beam aligned along the [100] zone axis to directly visualize the atomic positions within the film.

\bmhead{Transport measurements}
Wire connections for electric transport measurements were made using aluminum ultrasonic wire bonding. Temperature-dependent resistance and Hall effect measurements were performed using a Quantum Design Physical Property Measurement System (PPMS) at temperatures down to 2~K and magnetic fields up to 9~T. The excitation current was set to 10~$\mu$A to ensure high measurement accuracy. The Meissner state of the superconducting thin films was probed via direct-current magnetic susceptibility measurements using a Quantum Design Superconducting Quantum Interference Device (SQUID) magnetometer.

\bmhead{Acknowledgements}
We thank A.K. Cheetham for helpful discussions. We also acknowledge technical support from N.H. Teo, L. Jian, Y.H. Yin, J.Z. Li and Z.H. Yang. This work was supported by the National Research Foundation (NRF) Singapore under the NRF Investigatorship program and the Ministry of Education (MOE), Singapore, under its Tier-2 Academic Research Fund (AcRF), Grants No. MOE-T2EP50123-0013 and MOE-T2EP50124-0003.

\bmhead{Author contributions}
A.A. conceived and supervised the project; B.J., S.P., and A.A. planned and designed the experiments; B.J. prepared the samples with assistance from S.P., X.G., J.L., N.F., S.L. and T.A.; B.J. conducted the transport measurements with the help of S.P., K.Y.Y. and W.Z.; S.Z., Z.S.L and B.J. performed XRD and RSM measurements under the guidance of P.Y. and H.L.; Z.L., J-Y.C. and B.J. carried out STEM characterization and analysis with support from Q.H. and S.G.; B.J., S.P. and A.A. wrote the manuscript with contributions from all the authors.

\bmhead{Competing interests} The authors declare no competing interests.

\bmhead{Additional Information}
Correspondence and requests for materials should be addressed to Biemeng Jin or A. Ariando.

\newpage
\section*{Extended Data Figures}
\setcounter{figure}{0}
\renewcommand{\figurename}{{Extended Data Fig.}}
\renewcommand{\theHfigure}{supp.\arabic{figure}}

\begin{figure*}[h]
\centering
\includegraphics[width=0.9\columnwidth]{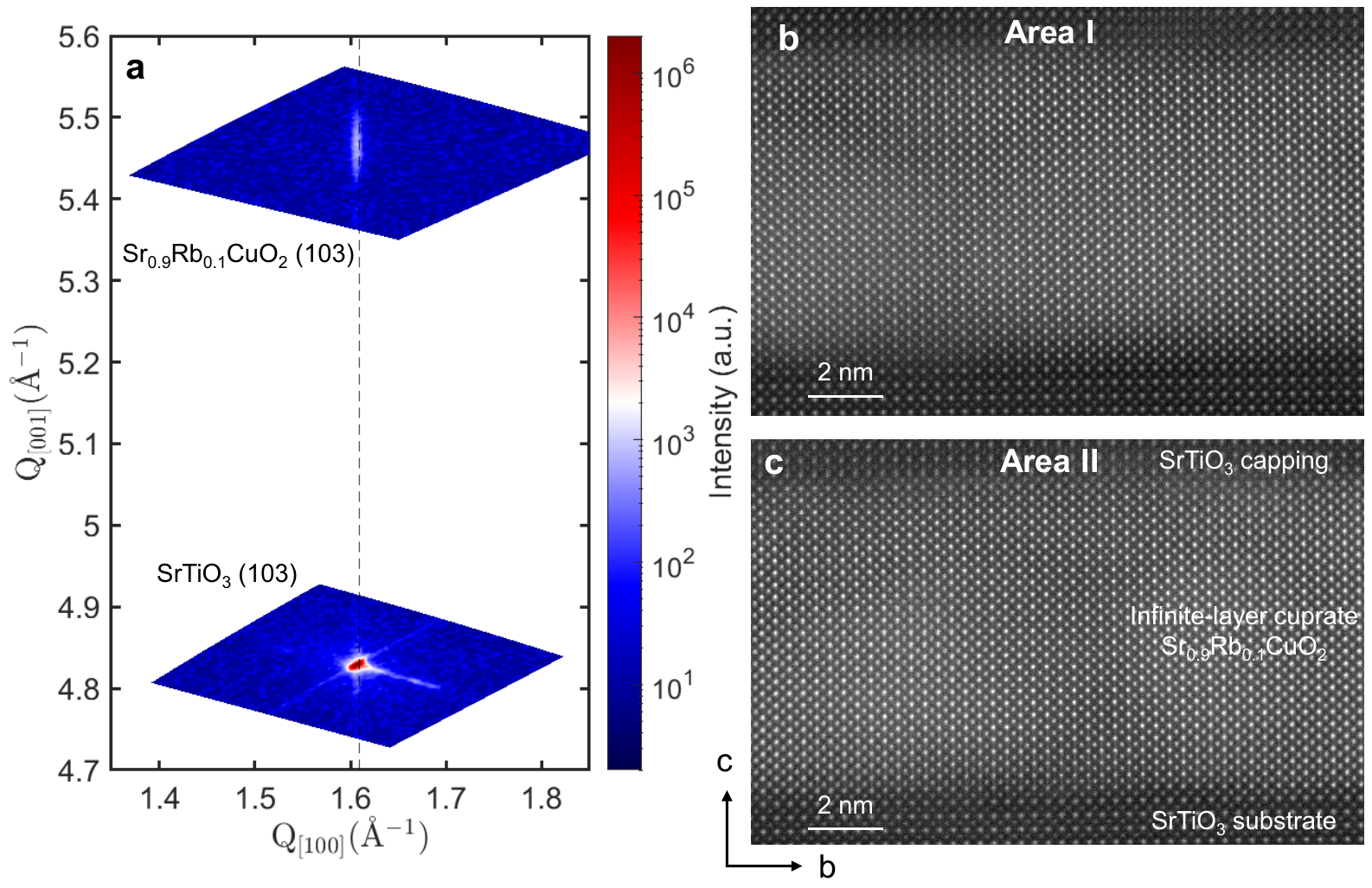}
\caption{
\textbf{a,} Reciprocal space mapping around the (103) peak of $\mathrm{Sr_{0.9}Rb_{0.1}CuO_2}$ thin films grown on $\mathrm{SrTiO_3}$ substrates, revealing an in-plane compressive strain of approximately 0.8\%.
\textbf{b–c,} Cross-sectional STEM–HAADF images of $\mathrm{Sr_{0.9}Rb_{0.1}CuO_2}$ thin films over large areas taken at different positions, showing high-quality infinite-layer lattices.
}\label{extendeddatafig1}
\end{figure*}

\begin{figure*}[h]
\centering
\includegraphics[width=0.9\columnwidth]{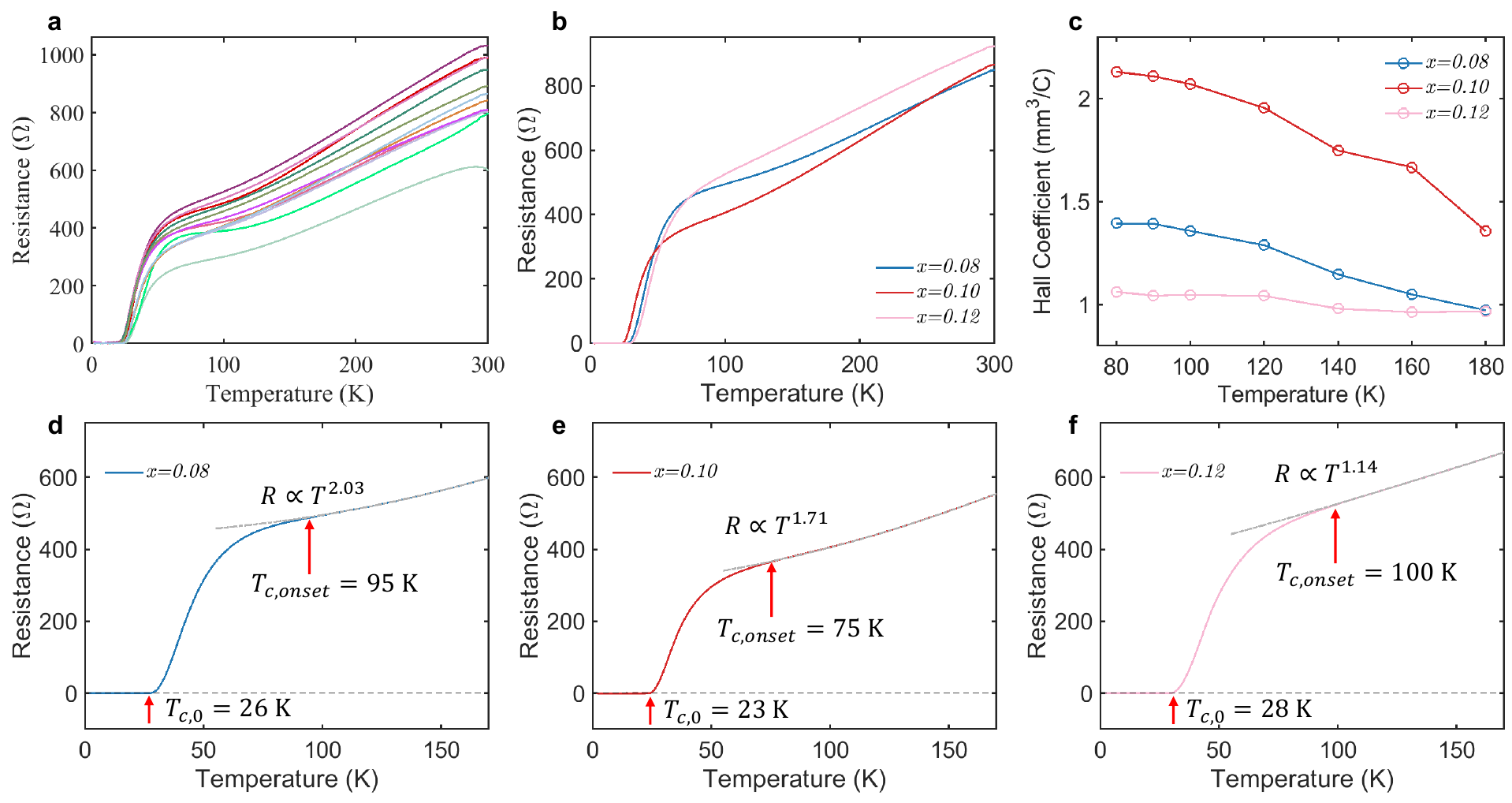}
\caption{
\textbf{a,} Temperature-dependent resistance $R(T)$ of multiple hole-doped infinite-layer cuprate $\mathrm{Sr_{0.9}Rb_{0.1}CuO_2}$ thin films, demonstrating the high reproducibility of the superconducting properties in these films.
\textbf{b,} Temperature-dependent resistance of hole-doped infinite-layer cuprates $\mathrm{Sr_{1-x}Rb_xCuO_2}$ with doping ratio $x = 0.08$, 0.10, and 0.12.
\textbf{c,} Temperature-dependent Hall coefficients of $\mathrm{Sr_{1-x}Rb_xCuO_2}$ with different $x =$ 0.08, 0.10, and 0.12. 
\textbf{d-f,} Determination of critical temperatures from fits to the 110-170~K range of $R$–$T$ curves for hole-doped superconducting $\mathrm{Sr_{1-x}Rb_xCuO_2}$ infinite-layer cuprate thin films with different doping ratios: (d) $x=0.08$, (e) $x=0.10$, and (f) $x=0.12$.
}\label{extendeddatafig2}
\end{figure*}

\begin{figure*}[h]
\centering
\includegraphics[width=0.9\columnwidth]{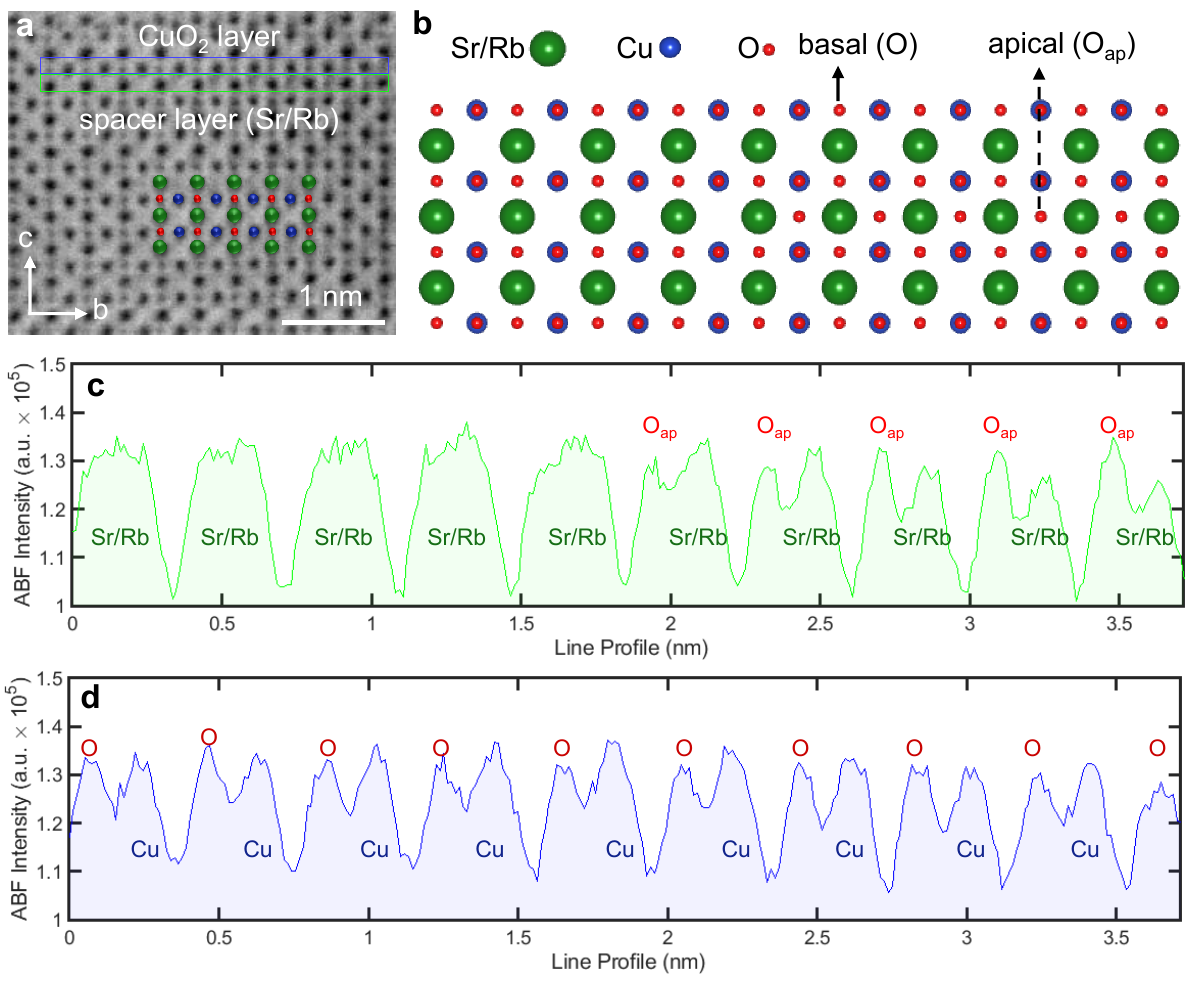}
\caption{
Visualization of apical oxygen in $\mathrm{Sr_{0.9}Rb_{0.1}CuO_2}$ infinite-layer cuprate by STEM–ABF imaging.
\textbf{a,} Cross-sectional STEM annular bright-field (ABF) image of a thin film $\mathrm{Sr_{0.9}Rb_{0.1}CuO_2}$, revealing the presence of apical oxygen in the spacer layer (green box). Dark features correspond to atomic columns.
\textbf{b,} Supercell structural model of $\mathrm{Sr_{1-x}Rb_xCuO_{2+\delta}}$, illustrating the spatial distribution of apical oxygen ($\mathrm{O_{ap}}$) and basal ($\mathrm{O}$) oxygen sites within the lattice. 
\textbf{c,} Line profile across the spacer layer (green box in a). The doublet peak corresponds to the Sr (or Rb) atomic column and the adjacent apical oxygen ($\mathrm{O_{ap}}$), confirming oxygen incorporation at the apical site.
\textbf{d,} Line profile across the CuO$_2$ layer (blue box in a), showing the atomic columns of Cu and in-plane basal oxygen.
}\label{extendeddatafig3}
\end{figure*}

\end{document}